\documentstyle[12pt,epsf,world_sci]{article}
\begin{document}

\title{{\bf $\Upsilon$ Production at CDF}}
\author{VAIA PAPADIMITRIOU\thanks{Representing the CDF Collaboration.}\\
{\em Fermi National Accelerator Laboratory, PO Box 500, Batavia,\\
Illinois 60510, U.S.A.}}
\vspace{0.3cm}

\maketitle
\setlength{\baselineskip}{2.6ex}

\begin{center}
\parbox{13.0cm}
{\begin{center} ABSTRACT \end{center}
{\small \hspace*{0.3cm} We report on measurements of the $\Upsilon$(1S),
$\Upsilon$(2S) and
$\Upsilon$(3S) differential and integrated cross sections
in $p \overline{p}$ collisions at $\sqrt{s} = 1.8$ TeV. The three
resonances were reconstucted through the decay
$\Upsilon \rightarrow \mu^{+} \mu^{-}.$  The cross section measurements
are compared to theoretical models of direct bottomonium production.}}
\end{center}

\section{Introduction}
We report a study of the reaction $p\bar{p}$ $\rightarrow$ $\Upsilon X$
$\rightarrow \mu^+ \mu^- X$ at \mbox{$\sqrt{s} = 1.8$ TeV}.
This study yields the $P_T$ dependence of the production
cross sections for the $\Upsilon$(1S), $\Upsilon$(2S),
and $\Upsilon$(3S) states and also the integrated over $P_t$ cross sections.
This is the first measurement of the individual $\Upsilon$ cross sections at
the Tevatron
energies and it is important for the investigation of bottomonium production
mechanisms in $p\bar{p}$ collisions~\cite{{1},{2}}.
Although, due to triggering constraints, we cannot extend our charmonia
production cross section measurements below 4 GeV/c, with the $\Upsilon$'s we
can go as low as $P_t = 0$, and therefore, besides the differential cross
sections, we can measure the total cross sections as well.

The CDF detector has been described in detail elsewhere~\cite{3}.
Here we give a brief description of the components relevant to this analysis.
The central tracking chamber
(CTC) is in a 1.4116--T axial magnetic field and has a
resolution
of $\delta P_t/ P_t = \sqrt{(0.0011P_T)^2 + (0.0066)^2}$ for beam
constrained tracks, where $P_t$ is the momentum transverse to the
beam direction.
The original central muon chambers (CMU), at a radius of 3.5~m
from the beam axis, provide muon identification in the region of
pseudorapidity \mbox{$|\eta^\mu| < 0.61$}. These chambers have been
complemented by the
addition of four layers of drift tubes behind 2 feet of steel (CMP), and as
a result, hadronic punch-through backgrounds to the muon signal have been
considerably reduced.

\section{Event Selection}
The measurements reported here are based on a data sample of opposite sign
dimuons collected with a multilevel trigger and with total integrated
luminosity of 16.6~$\pm$~0.6~pb$^{-1}$.
The muons from the $\Upsilon \rightarrow \mu^+\mu^-$ decay were required to
satisfy the following criteria: Both muons are identified by the CMU system and
at least one of them is identified by both the CMU and CMP systems;
$P_t$($\mu$) $> $ 2.0 GeV/c for each muon;
$P_t$($\mu$) $> $ 2.8 GeV/c for at least one of the two muons;
less than a $(3-4) \sigma$ difference in position
between each muon chamber track and its associated, extrapolated CTC track,
where $\sigma$ is the calculated uncertainty due to
multiple scattering, energy loss, and measurement uncertainties;
a common vertex along the beam axis for the two muons; region of rapidity
of the dimuon pair $|y| < 0.4$.
In addition the Level 1, Level 2 trigger had to be satisfied by the dimuon
pair.
The transverse momentum $P_t$ of the dimuon pairs had
to be greater than 0.5 GeV/c and less than 20 GeV/c.
The invariant mass distribution of opposite sign dimuon pairs
is shown in Fig. 1.
\begin{figure}[htbp]
\footnotesize
\begin{center}
        \mbox{
\epsfxsize=3.5in
\epsfysize=3.5in
\epsffile {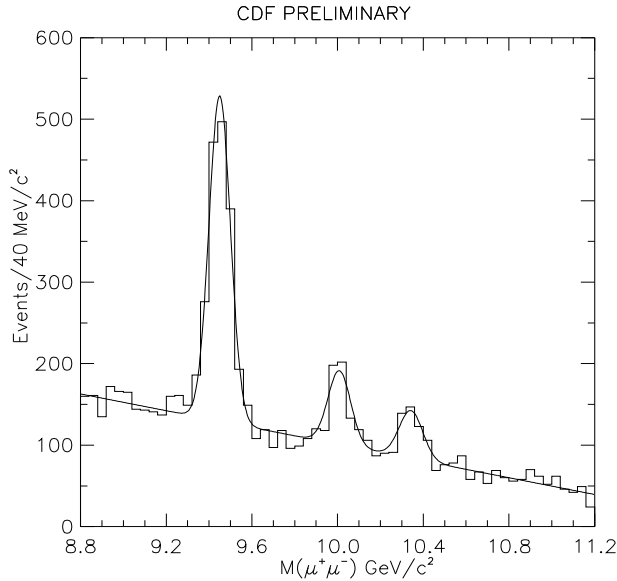}
}
\vspace*{0.05in}
\caption{{\label{fig:mass}}
{Invariant mass distribution for $\Upsilon$(1S), $\Upsilon$(2S) and
$\Upsilon$(3S).}}
\label{fig:mass}
\end{center}
\end{figure}

\vspace{0.2cm}
\section{Acceptance and other Efficiencies}
The geometric and kinematic acceptances for $\Upsilon$(1S),$\Upsilon$(2S),
$\Upsilon$(3S) $\rightarrow \mu^+\mu^-$ were evaluated by a dimuon event
generator that produces $\Upsilon$'s with flat $P_t$ and $y$ distributions.
The generated events were processed with the full detector simulation and
the same reconstruction used for the data. The acceptances are $P_t$ and
$y$ dependent and they are very similar for the three different states.
To verify that the acceptance is independent of the kinematic distribution
of generated events, the acceptance calculation was repeated for the
$\Upsilon$(1S) state using a parton level generator~\cite{1}
which provides on an event by event basis the four momentum of various
bottomonium states decaying to $\Upsilon$(1S).
This generator together with the CLEO decay table was used to obtain
the theoretical predictions shown in Figures 2 and 3.

Efficiency corrections required for the $\Upsilon$ cross section
calculations are as following:
the Level 1 and Level 2 trigger efficiencies for each muon are
increasing functions
of $P_t$, reaching a plateau of $92 \%$ above 3.1 GeV/$c$.
The level 3 tracking efficiency is $92\pm2\%$.
The total trigger efficiency for each dimuon event
is taken to be the
product of the Level 1 and
Level 2 efficiencies for each muon and the level 3 efficiency for the
event.
The offline CTC track reconstruction efficiency is (97.8 $\pm$ 1.4)\%
and the muon reconstruction efficiency is $(96\pm1.4)\%$ for each dimuon
event.
The matching cut on the difference between the muon chamber track
and the extrapolated CTC track is $(98.7\pm0.2)\%$ efficient.
\vspace{0.3cm}
\section{Differential and total cross sections}
The acceptance and efficiency corrected $\Upsilon$ cross sections are
displayed in Figures 2 and 3 as functions of $P_t$.
The vertical error bars are from statistical fluctuations
in the number of counts (background fluctuations included) and the
$P_t$--dependent systematic uncertainties added in quadrature.
The signal and background contributions have been determined independently
in the $P_t$ bins shown in the differential cross section plots,
by fitting the dimuon mass distribution in each bin to a gaussian plus a
first degree polynomial.
\begin{figure}[htbp]
\footnotesize
\begin{center}
        \mbox{
\epsfxsize=2.9in
\epsffile {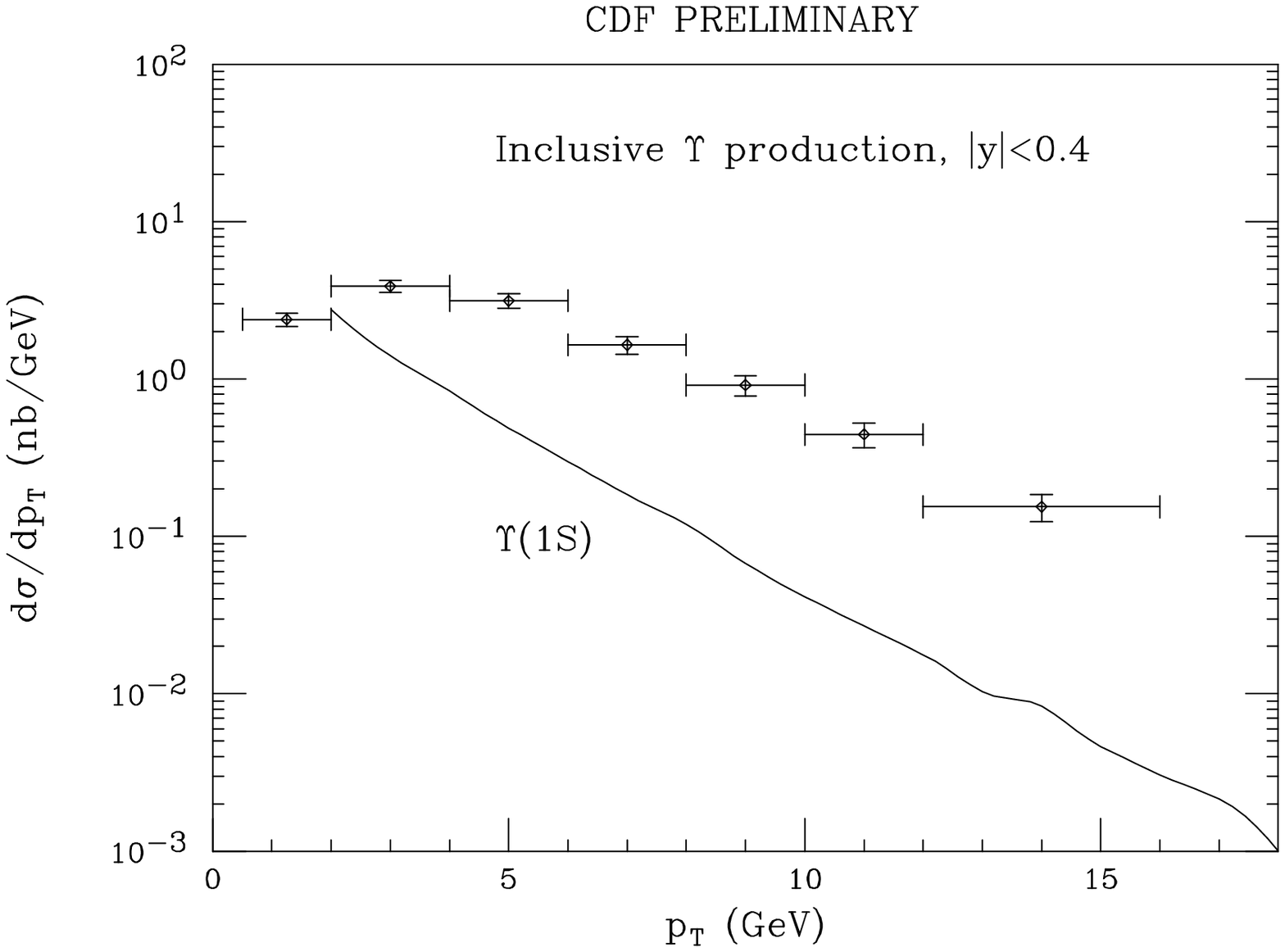}
}
\hfill
        \mbox{
\epsfxsize=2.9in
\epsffile {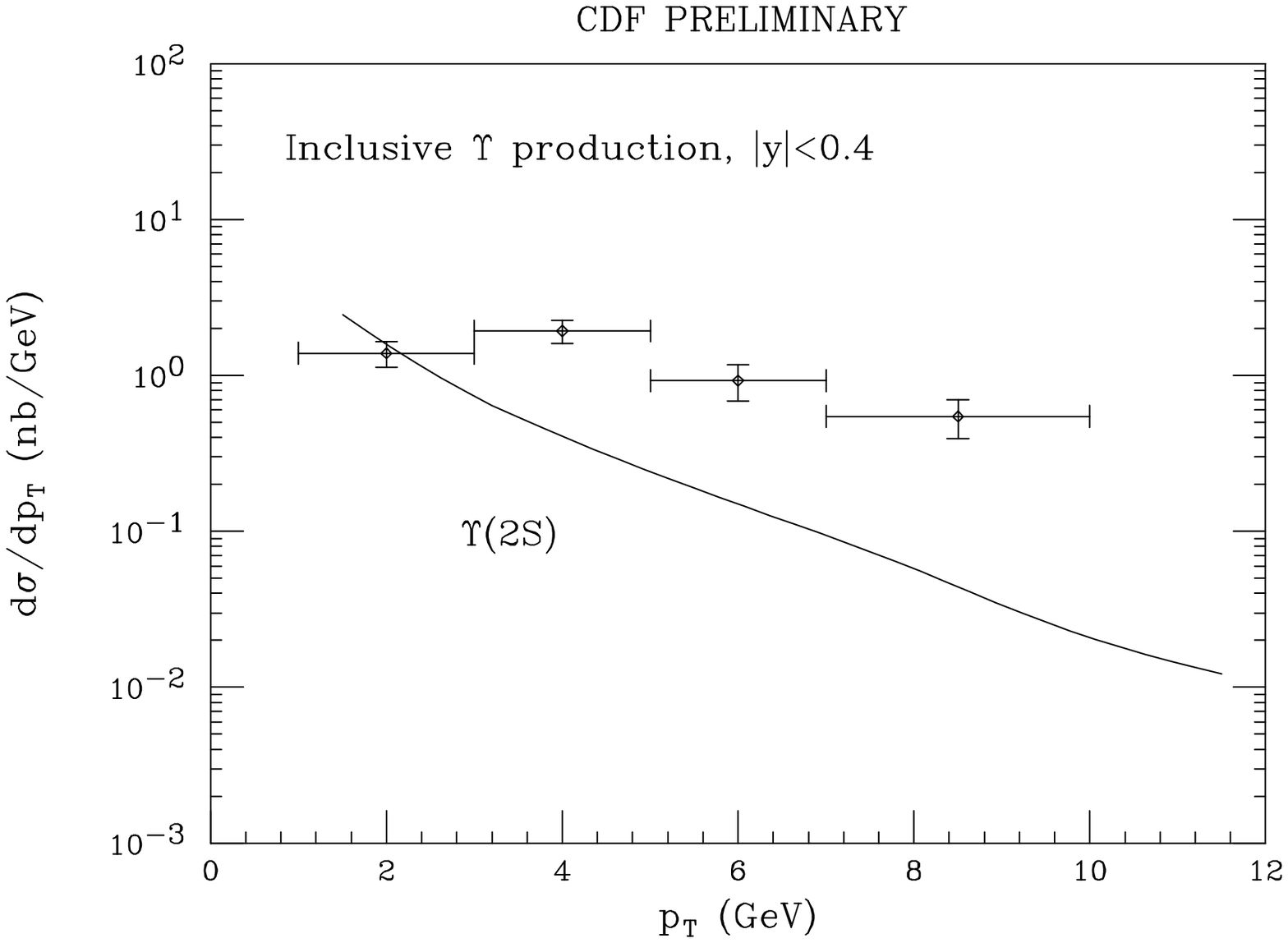}
}
\vspace*{0.10in}
\caption{{\label{fig:up1s}}
$\Upsilon$(1S) and $\Upsilon$(2S) differential cross section
for $|y| <$ 0.4
compared to the theoretical prediction. Error bars include
statistical error and $P_T$ dependent systematic error. There is also a common
systematic error of 15\%(22\%) for the $\Upsilon$(1S)($\Upsilon$(2S))
 not shown.The theoretical curves are leading order and
were generated using MRSDO PDF with scale
$\mu^2 = P_t^2 + m_\Upsilon^2$. Production and decay of higher bottomonium
states are included.}
\label{fig:Up1s}
\end{center}
\end{figure}

\begin{figure}[htbp]
\footnotesize
\begin{center}
        \mbox{
\epsfxsize=3.7in
\epsfxsize=3.7in
\epsffile {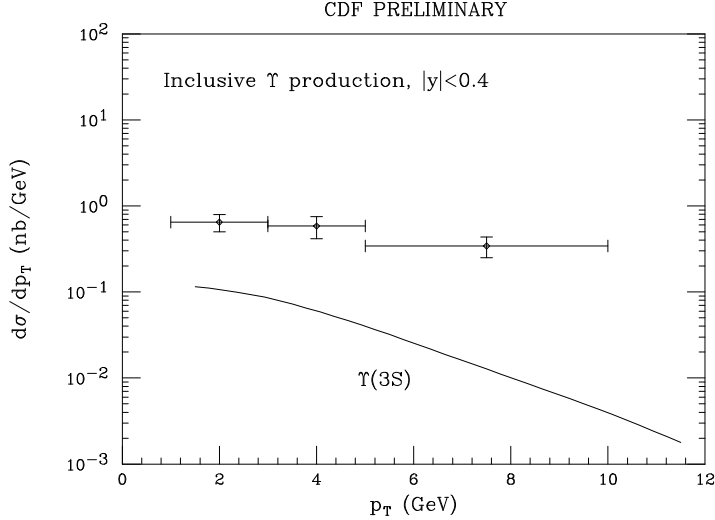}
}
\vspace*{0.10in}
\caption{{\label{fig:up3s}}
$\Upsilon$(3S) differential cross section for $|y| <$ 0.4
compared to the theoretical prediction. Error bars include
statistical error and $P_T$ dependent systematic error. There is also a common
systematic error of 18\%
 not shown.The theoretical curves are leading order and
were generated using MRSDO PDF with scale
$\mu^2 = P_t^2 + m_\Upsilon^2$.}
\label{fig:Up3s}
\end{center}
\end{figure}

The integrated cross section results are:
\begin{eqnarray*}
\sigma(\overline{p} p \rightarrow \Upsilon{\rm (1S)}, |y| < 0.4,
P_t > 0.5 {\rm \: GeV/c}) & = &
23.48 \pm 0.99\: {\rm (stat)} \pm 2.80\: {\rm (sys)\: nb} \\
\sigma(\overline{p} p \rightarrow \Upsilon{\rm (2S)}, |y| < 0.4,
P_t > 1.0 {\rm \: GeV/c}) & = &
\: \: 10.07 \pm 1.01\: {\rm (stat)} \pm 1.99\: {\rm (sys)\: nb} \\
\sigma(\overline{p} p \rightarrow \Upsilon{\rm (3S)}, |y| < 0.4,
P_t > 1.0 {\rm \: GeV/c}) & = &
\: \: 4.79 \pm 0.64\: {\rm (stat)} \pm 0.72\: {\rm (sys)\: nb.} \\
\end{eqnarray*}
The systematic errors for the total cross section measurements are coming
from the trigger efficiency (4-5\%), from the acceptance model (10\%),
from the luminosity determination (3.6\%) and from the various
reconstruction
efficiencies (2.8\%). We have an additional systematic coming from the
uncertainty of the branching ratios of $\Upsilon \rightarrow
\mu^{+} \mu^{-}$, which is 2.4\%, 16\% and 9.4\% for the $\Upsilon$(1S),
$\Upsilon$(2S) and $\Upsilon$(3S) respectively.
Total cross sections in the range $|y| <$0.4 can be calculated at LO
QCD~\cite{2}. Use of MRSD0 PDF with scale $\mu^2 = m_\Upsilon^2$ yields
4.2, 2.5 and 0.12 nb for the $\Upsilon$(1S),
$\Upsilon$(2S) and $\Upsilon$(3S) respectively~\cite{1}. These results
include the production and decay of $\chi_b(1P)$ and $\chi_b(2P)$ states,
which are found to dominate the rate of $\Upsilon$(1S) and $\Upsilon$(2S)
respectively. No $\chi_b(3P)$ state has been observed at present, and
therefore its possible contribution to $\Upsilon$(3S) is not included.
Our measurements are a factor of 4-6 higher than the theoretical
prediction for the $\Upsilon$(1S) and
$\Upsilon$(2S) states, and by a factor of 40 for the $\Upsilon$(3S) state.
While a factor 4-6 can be partly accomodated by the inclusion of higher
order corrections and possible new production mechanisms~\cite{4},
the big discrepancy for the $\Upsilon$(3S) state suggests that there are
additional $\chi_b$ states below the $B\bar B$ threshold that contribute
significantly to the $\Upsilon$(3S) production.


\bibliographystyle{unsrt}

\end{document}